\documentclass[twocolumn,preprintnumbers,amsmath,amssymb,prb]{revtex4}

\usepackage{graphicx}
\usepackage{dcolumn}
\usepackage{bm}

\begin{document}
\title{Gas-solid coexistence of adhesive spheres}
\author{P. Charbonneau}
\affiliation{FOM Institute for Atomic and Molecular Physics,
Kruislaan 407, 1098 SJ Amsterdam, The Netherlands}
\author{D. Frenkel} \email{frenkel@amolf.nl}
\affiliation{FOM Institute for Atomic and Molecular Physics,
Kruislaan 407, 1098 SJ Amsterdam, The Netherlands}



\date{\today}
\maketitle

The Baxter limit of adhesive spheres~\cite{baxter:1968} is
often deemed an unphysical model for liquids, due to the
thermodynamic metastability of its disordered phases with
respect to crystal formation~\cite{stell:1991,sear:1999b}. Here we
suggest conditions under which the disordered phases are
actually stable with respect to the solid for narrow attractive
interactions down to the Baxter limit. Furthermore, using basic
free energy considerations we estimate the necessary condition
for the gas-liquid critical point to be stable with respect to
crystallization~\cite{miller:2003}. Possible experimental and
simulation realizations are briefly discussed.

For what follows, it is convenient to consider the original
Baxter adhesive sphere model, i.e. a square-well fluid with
well depth $\epsilon$ and well width $\delta$, in hard-core
diameter $\sigma$ units, which we set equal to unity for
convenience. Except where explicitly mentioned, the results are
for three-dimensional systems. For adhesive spheres, the
stickiness $\tau$ behaves as a temperature-like quantity, which
is defined as
\begin{equation}
\tau^{-1}\equiv 12\delta e^{\epsilon/k_BT}, \label{eq:taudef}
\end{equation}
where $k_BT$ is the product of Boltzmann's constant with
temperature. The equivalent quantity for square-well potentials with finite
width~\cite{miller:2004} is
\begin{equation}
\tau^{-1}=4\left(\left(1+\delta\right)^3-1\right)\left(e^{\epsilon/k_BT}-1\right).
\label{eq:tausw}
\end{equation}
The second virial coefficient can
then be expressed as $B_{2v}/B_{2v}^{HS}=1-\frac{1}{4\tau}$,
where $B_{2v}^{HS}=2\pi/3$ is the hard-sphere value.

The critical point of sticky spheres (and, to a good
approximation, {\em all} fluids with a narrow, isotropic
attraction~\cite{noro:2000}) occurs at $\tau_c\approx
0.11$~\cite{miller:2003}. How about freezing? At low
temperatures, the density of the gas coexisting with the solid
is very low, $\rho_g\ll1$, while at high temperatures, the
freezing density is that of hard spheres. Typically, the
freezing curve moves rather abruptly, but continuously, from
one limit to the other. We estimate the temperature and
corresponding $\tau$ where this happens and compare it with the
known $\tau_c$~\cite{miller:2003}.

\begin{figure}
\center{\includegraphics[scale=.25]{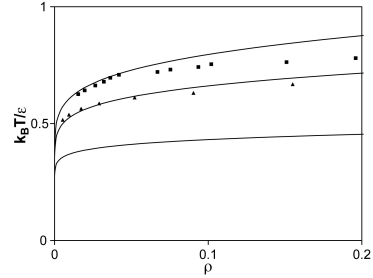}}\\
\caption
{Gas-solid coexistence line from
Eq.~\ref{eq:swcoex} for square-well interaction with $Z=12$ and
$\delta=0.25$, $0.15$, and $0.03$, from top to bottom.
Simulation results for the first two cases are included for
comparison~\cite{pagan:2005,liu:2005}.} \label{fig:pdrhoT}
\end{figure}

\begin{figure}
\center{\includegraphics[scale=.25]{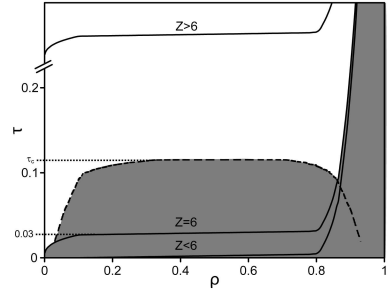}}\\
\caption
{Sketch of the three-dimensional
Baxter phase diagram following ref.~\cite{miller:2003} for the
isotropic gas-liquid binodal (dashed line) and this work for
the gas-solid coexistence curves (solid lines). The gas-solid
coexistence is found at high $\tau$ for $Z>6$ and at very low
$\tau$ for $Z<6$, while it follows Eq.~\ref{eq:baxtercoex} for
$Z=6$. Gas-crystal coexistence curves for $\rho\gtrsim0.1$ and
gas-liquid coexistence for $\tau<0.1$ are drawn for schematic
purposes only.} \label{fig:BaxterPD}
\end{figure}

To estimate the crossover temperature where the density of the
vapor coexisting with the solid starts to increase, we make use
of the fact that the chemical potential per particle $\mu$ is
the same in the different coexisting phases. For the fluid, we
use an ideal gas (we check this approximation later), while we
treat the crystal as a collection of uncorrelated cells each
with free volume $v_f$. Although this last approximation is
rather crude compared to what is usually used in similar
systems~\cite{tejero:1995}, for narrow square-well potentials
at low temperatures, it is not unreasonable. This gives
\begin{eqnarray}
\frac{\Delta \mu}{k_BT}&=&\frac{\Delta e}{k_BT}-\frac{\Delta s}{k_B}+\frac{p \Delta v}{k_BT}\nonumber\\
&\approx& -\frac{Z\epsilon}{2k_BT}-\ln{\rho_g v_f}-1,
\label{eq:eqrelation}
\end{eqnarray}
where $s$ is the entropy, $v$ is the volume, $e$ is the energy,
and $Z$ is the coordination number of the crystal. To obtain
the last line, we also assume $v$ to be much smaller in the
crystal than in the gas, so $\Delta v\approx \rho_{g}^{-1}$.
Note that when $\Delta \mu=0$, $k_BT_{coex}/Z\epsilon$ is the
only temperature term, thus changing the coordination number of
the solid effectively rescales the coexistence temperature. In
other words, to a first approximation, the gas coexistence
curve for a solid with coordination number $Z_2$ can be
obtained from the known gas coexistence curve for a solid with
coordination number $Z_1$ by multiplying $T_{coex}$ by
$Z_2/Z_1$.

Following Sear~\cite{sear:1999b}, we take
$v_f\approx\left(\frac{\delta}{2}\right)^d$ for narrow square
wells in $d$ dimensions. Eq.~\ref{eq:eqrelation} can then be
used to calculate gas-solid phase coexistence directly,
\begin{equation}
\frac{k_BT_{coex}}{\epsilon}=\frac{-Z/2}{\ln{\rho_g\left(\frac{\delta}{2}\right)^d}+1}.
\label{eq:swcoex}
\end{equation}
This is compared to published numerical results of $Z=12$
square-well phase diagrams in Fig.~\ref{fig:pdrhoT}. At low
$\rho_g$, towards the wider limit of narrow wells, there is a
fairly good agreement between this crude treatment and the MC
simulated phase diagrams. Results for $Z=4$ also qualitatively
agree with this picture~\cite{zhang:2005}.

As a last consistency check, we examine in what regime
corrections to $\Delta v\approx\rho_g^{-1}$ are small. The next
order terms are the crystal's finite volume, which is of ${\cal
O}(1)$, and the fluid's second-virial coefficient $B_{2v}$.
Using Eq.~\ref{eq:swcoex} as a mean-field-like closure for
$\epsilon/T_{coex}$ gives
\begin{eqnarray}
\frac{B_{2v}^{coex}}{B_2^{HS}}&\approx&1-3\delta\left(e^{\epsilon/T_{coex}}-1\right)\nonumber\\
&\approx&1-3\delta\left(\left(\frac{8}{e\delta^3\rho_g}\right)^{2/Z}-1\right),
\end{eqnarray}
which for $Z\geq6$ leaves an ${\cal O}(\rho_g^{-2/Z})$
correction. At low densities the two contributing factors are
both $\ll \rho_g^{-1}$ and have an opposite sign, so they
amount to an error in $T_{coex}$ of less than a few percents
for narrow well widths when $\rho_g\lesssim0.1$. This level of
precision is sufficient for the sake of our argument.

Going to the Baxter limit, $\delta\rightarrow 0$, we then use
the definition of stickiness from Eq.~\ref{eq:taudef} to
rewrite Eq.~\ref{eq:swcoex} as
\begin{equation}
1 = -\ln
\left[2^{-d}\rho_g\delta^{d-Z/2}\left(\frac{1}{12\tau_{coex}}\right)^{Z/2}\right].
\label{eq:baxterequil}
\end{equation}
For $Z>2d$, $\delta^{d-Z/2}$ diverges, while at freezing the
RHS of Eq.~\ref{eq:baxterequil} should be ${\cal O}(1)$. If we
fix the density of the gas at a small but finite value, it
follows that $\tau_{coex}\rightarrow\infty$, i.e. the fluid
freezes well above $\tau_c\approx0.11$. Conversely, when
$Z<2d$, $\tau_{coex}=0$, so the triple point also lies on the
$\tau=0$ axis. The interesting case is the marginal situation
$Z=2d$, where we obtain a non-trivial fixed crossover
$\tau_{coex}$. For $d=3$, the crossover $\tau_{coex}$ increases
by less than $15\%$ as $\delta$ varies from $0.25$ down to the
Baxter limit, which is small relative to the divergence
obtained for $Z>2d$. Moreover we expect $\tau_{coex}$ to lie in
the same regime as $\tau_c$. Indeed, in three dimensions
Eq.~\ref{eq:baxterequil} reduces to
\begin{equation}
\tau_{coex}=\frac{\left(e\rho_g\right)^{1/3}}{24}.
\label{eq:baxtercoex}
\end{equation}
Using $\rho_g=0.1$ as a reference point, we then obtain a
crossover $\tau_{coex}\approx0.03$, which is well below
$\tau_c$ for the isotropic case, as shown in
Fig.~\ref{fig:BaxterPD}. The gas-liquid critical point is thus
{\em stable} with respect to the crystal.

Simple cubic crystals, among others, and isostatic solids,
fulfill the $Z=2d$ condition for all $d$. On the other hand,
monodisperse, isotropically adhesive spheres, have invariably
$Z=12$ crystal phases. However, if either the spheres are not
monodisperse or if the interaction is patchy, a lower $Z$ can
be obtained. In the adhesive limit, even infinitesimally
polydisperse solids eventually fractionate~\cite{sear:1999b},
reaching thermodynamic equilibrium, which then results in
$Z\approx12$. However, this is kinetically nearly impossible
and therefore of little experimental relevance. Instead, an
isostatic polydisperse metastable crystal would leave $\tau_c$
stable with respect to the gas-solid coexistence. For patchy
adhesive spheres $Z$ is precisely controlled by the number of
patches. Though $\tau_c$ also varies with
$Z$~\cite{bianchi:2006}, this effect is unlikely to be
sufficient to bring it below $\tau_{coex}$.

In conclusion, if one manages to reduce the bonding of the
solid phase down to $Z\leq2d$, the high density fluid remains
stable with respect to the solid down to the Baxter limit.
Based on rough, but reasonable estimates, the three-dimensional
gas-liquid critical point would then be also stable with
respect to crystallization. Furthermore, this scheme might be
realizable both experimentally and through simulation.

\begin{acknowledgments}
PC would like to thank D.R. Reichman for insightful
discussions. PC also acknowledges Marie-Curie IIF XXXXXXXXX funding.
This work is part of the research program of the ``Stichting voor
Fundamenteel Onderzoek der Materie (FOM)'', which is financially
supported by the ``Nederlandse organisatie voor Wetenschappelijk Onderzoek (NWO)''.
\end{acknowledgments}

\bibliography{PhaseDiag}
\end{document}